\documentclass[a4paper,12pt]{article}

\usepackage{graphicx}
\usepackage{amssymb}
\usepackage{cite, epsfig}
\usepackage{curves}

\def\be{\begin{equation}} 
\def\ee{\end{equation}} 
\def\bea{\begin{eqnarray}} 
\def\eea{\end{eqnarray}} 
\def\beb{\begin{eqnarray*}}
\def\eeb{\end{eqnarray*}}

\newcommand{\qom}[1]{{\omega_{\vec {#1}}}}

\begin{document}
\makeatletter
\def\fmslash{\@ifnextchar[{\fmsl@sh}{\fmsl@sh[0mu]}}
\def\fmsl@sh[#1]#2{%
  \mathchoice
    {\@fmsl@sh\displaystyle{#1}{#2}}%
    {\@fmsl@sh\textstyle{#1}{#2}}%
    {\@fmsl@sh\scriptstyle{#1}{#2}}%
    {\@fmsl@sh\scriptscriptstyle{#1}{#2}}}
\def\@fmsl@sh#1#2#3{\m@th\ooalign{$\hfil#1\mkern#2/\hfil$\crcr$#1#3$}}
\makeatother
\thispagestyle{empty}
\begin{titlepage}

\begin{flushright}
TUW-04-03
\end{flushright}

\vspace{0.3cm}
\boldmath
\begin{center}
  {\Large {\bf 
  Towards UV Finite Quantum Field Theories from Non-Local Field Operators
   }}
\end{center}
\unboldmath
\vspace{0.8cm}
\begin{center}
{
{{\bf
Stefan~Denk\footnote{denk@hep.itp.tuwien.ac.at}, 
Volkmar~Putz\footnote{putz@hep.itp.tuwien.ac.at}, 
Manfred~Schweda\footnote{mschweda@tph.tuwien.ac.at} and 
Michael~Wohlgenannt\footnote{miw@hep.itp.tuwien.ac.at}
}}}
\end{center}
\vskip 1em
\begin{center}

Institut f\"ur theoretische Physik, Technische Universit\"at Wien,\\
Wiedner Hauptstra{\ss}e 8 - 10, A - 1040 Wien, Austria\\

 \end{center}

\vspace{\fill}

\begin{abstract}
\noindent
A non-local toy model whose interaction consists of smeared,
non-local field operators is presented. We work out the Feynman rules and
propose a power counting formula for arbitrary graphs. Explicit calculations for
one loop graphs show that their contribution is finite for sufficient smearing 
and agree with the power counting formula. UV/IR mixing does not
occur. 
\end{abstract}

\end{titlepage}

\section{\label{sec1}Introduction}



Already in \cite{Pais:1950za} the behaviour of non-local field theories has
been studied, and it has been questioned whether they help to avoid
divergences. In a different approach, \cite{Ohanian:1997ni}, the 
construction of finite field theories starting from smeared propagators has been
investigated. The smeared propagators are considered as a result of
gravitational fluctuations at the Planck scale.
We want to follow this line of thought and restrict ourselves to scalar field
theory. But our starting point is a non-local
deformation of the field operators in the interaction, leaving the free
Hamiltonian untouched.

Beside the above mentioned approaches to non-local field theories, we want to 
address four other ones and to distinguish them clearly from our point of 
view presented here.

In the first approach \cite{Filk:1996dm}, the point-wise multiplication of 
scalar fields in the Lagrangian is replaced by a non-local $*$-product. 
The $*$-product is such that quadratic terms are unaltered,
$$
\int d^4 x \, A*B = \int d^4 x\, AB.
$$
The Feynman rules are obtained directly from the classical action.
Therefore, propagators are unchanged. The only modifications are due to the 
vertex contributions,
\be
\Gamma_4 = \int d^4 x \, (\phi * \phi * \phi * \phi) (x).
\ee 
Each vertex contributes, beside the coupling constant,
a phase factor. In momentum space we get
$$
\tilde V (k_1,\dots, k_n) = \delta(k_1+\dots+k_n)\, \exp
        ( -{i\over 2}\sum_{i<j}^n {k_i}_\mu {k_j}_\nu \sigma^{\mu\nu} ),
$$
where $k_1,\dots, k_n$ are the incoming momenta, and $\sigma^{\mu\nu}$ is the 
real antisymmetric deformation parameter of dimension $[length]^2$
$$
{}[q^\mu,q^\nu] = i \, \sigma^{\mu\nu}.
$$
The drawbacks of this theory are the so-called UV/IR mixing \cite{uv-ir} and its non-unitarity 
\cite{Gomis:2000zz}.

Unitarity can be restored by considering the Hamiltonian instead of the Lagrangian and by computing
Feynman rules using the Gell-Mann-Low formula (\ref{gell-mann-low}) \cite{Bahns:2002vm,
Liao:2002xc, Liao:2002pj, Bozkaya:2002at, Denk:2003jj,Fischer:2003jh}. These methods represent a second possibility to describe
non-commutative quantum field theories perturbatively. 
The free propagator is unchanged due to the 
remarkable fact that the deformed free
Hamiltonian $H_0^*$ is equal to the undeformed one,
\bea
\nonumber
H_0^* & = & \int d^3 x\, \bigg(\sum_\mu \partial_\mu \phi * \partial_\mu \phi + m^2 \phi* \phi \bigg)
        = \int d^3 x\, \bigg(\sum_\mu \partial_\mu \phi \, \partial_\mu \phi + m^2 \phi \, \phi \bigg) \\
\label{H-def}
& = & H_0,
\eea
with $\phi$ representing free field operators.
We will elaborate on this important statement in more detail and generality elsewhere.

The third approach is based on an oscillator representation of non-commu\-tative 
space-time \cite{Chaichian:1998kp,Cho:1999sg,Smailagic:2003rp}.
Let us focus on the presentation given in 
\cite{Smailagic:2003rp}. Scalar field theory in $D=2+1$ dimensions is
considered. The time component seems artificial. In this sense, the results
obtained in \cite{Chaichian:1998kp,Cho:1999sg} for $4$ dimensional Euclidean
space agree, corresponding to $D=4+1$ dimensional Minkowski space according to 
\cite{Smailagic:2003rp}.

In \cite{Smailagic:2003rp}, time
commutes with the spatial coordinates which satisfy the relation
\be\label{b1}
{}[\hat x^i, \hat x^j ] = i \theta \epsilon^{ij},
\ee
$i,j=1,2$. Further on, there are the usual commutation relations with the 
momenta,
\be\label{b2}
{}[\hat x^i, \hat p_j ] = i \delta^i _j\, , \qquad
[\hat p_i, \hat p_j ] = 0.
\ee
New coordinates $\hat z$ and $\hat z^\dagger$ are introduced \cite{Glauber:1963tx},
\bea
\label{b3}
\hat z & = & {1\over \sqrt{2} } (\hat x^1 + i \hat x^2 ),\\
\nonumber
\hat z^\dagger & = & {1\over \sqrt{2} } (\hat x^1 - i \hat x^2 )
\eea
in order to obtain
\be
\label{b4}
{}[\hat z , \hat z^\dagger ] = \theta. 
\ee
$\hat z$ and $\hat z^\dagger$ can be established as annihilation and creation operators of 
a harmonic oscillator, and coherent states can be used as a basis of the Fock space. 
Coherent states $|z>$ are eigenstates of the annihilation operator,
\be
\label{b5}
\hat z |z\rangle  = z |z\rangle, \quad \langle z| \hat z^\dagger =\bar z \langle z|.
\ee 
They are given by
\be
|z\rangle = \exp ( -{z\bar z\over 2 \theta} - {z\over \theta} \hat z^\dagger ) |0\rangle,
\ee
satisfying the completeness relation
\be
{1\over \pi\theta} \int dz\, d\bar z \, |z\rangle\langle z| = 1.
\ee
Coherent states are not orthogonal, however,
\be
\langle w| z \rangle = \exp ( -{|z|^2 + |w|^2 \over 2\theta } - {\bar w z \over  \theta} ).
\ee
Via expectation values, one can assign ordinary functions to any operator $F(\hat x^1, \hat x^2)$,
\be
F(z) := \langle z | F(\hat x^1, \hat x^2) | z \rangle.
\ee
The algebraic structure of the non-commutative algebra (\ref{b1}) is properly taken care of, i.e.
\be
\langle z | [\hat x^1, \hat x^2 ] | z \rangle = i\theta.
\ee
With the expansion of a real scalar free field operator
\bea
\phi(t,z) & = & \int {d^2p\over 2\pi}\, b_p \exp (-iEt) \langle z | \exp
(ip_j \hat x^j) | z \rangle + h.c.,\\
(\Box_x + m^2) \phi(t,x) & = & 0, \nonumber
\eea
the propagator --- defined as the expectation value of a time ordered product of field 
operators --- becomes
\bea
&& \hspace{-1.5cm} G(t_1-t_2, z_1-z_2) =  \langle 0 | T \phi(t_1,z_1)
	\phi(t_2,z_2) | 0 \rangle \\
\nonumber
&& = \int {dE \, d^2 p \over (2\pi)^{3/2}}  \,
	{-1\over E^2-\vec{p}\,{}^2-m^2 } \, \exp ( -{\theta\over 2} 
	\vec{p}\,{}^2 ) \exp(-i E(t_1-t_2)) \\
\nonumber
\label{it-damp}
&& \times\, \exp\left(i {p_1 \over \sqrt{2}}(z_1 - z_2 + 
	\bar z_1 - \bar z_2) + i {p_2 \over \sqrt{2}}
	(z_1 - z_2 - \bar z_1 + \bar z_2)\right).
\eea
This propagator is the "Green's function" of the ordinary Klein-Gordon equation, 
with the exception that the delta function is replaced by an approximate (smeared) delta 
function,
\bea
\nonumber
& (\Box_1+m^2) \, G(t_1-t_2, z_1-z_2) = \left( -\partial_{t_1}^2 + 2 \partial_{z_1}
        \partial_{\bar z_1} + m^2\right) G(t_1-t_2, z_1-z_2) & \\
& = {2\pi\delta(t_1-t_2)\over \theta} \, \exp\left( -{1\over 4\theta}(z_1 - z_2 +
\bar z_1 - \bar z_2)^2 + {1\over 4\theta}(z_1 - z_2 - \bar z_1 + \bar z_2)^2 \right).
&
\eea
In this case, the free propagator is modified. It experiences an exponential
damping (\ref{it-damp}). It is important to note that the non-commutativity is
related to exponentially damped propagators. This fact motivates our model.

In the fourth approach \cite{Bahns:2003vb}, also only the interaction
Hamiltonian is modified. The fields are smeared over space-time in the following
way
\bea
H^*_I(t) & = & \lambda c_n \int d^3 x \int_{\mathbb{R}^{4n}} da_1\cdots da_n 
        :\phi(x+a_1)\cdots \phi(x+a_n):\\
\nonumber
& \times & \exp\left( -{1\over 2}\sum_{j,\mu} {a_j^\mu}^2 \right) \delta^{(4)}
\left( {1\over n}\sum_{j=1}^n a_j \right). 
\eea
Using this ansatz, it has been shown that the Dyson expansion of the S-matrix is
finite, order by order.

Similar to the second and fourth approach above, we consider only modifications
in the interaction. We replace the local field operators $\phi$ by smeared,
non-local fields $\phi_M$, as discussed in the next Section. Therefore, the free
propagators are not modified. Internal lines, however, will be modified by an
exponential damping factor, similar to the third approach. Let us emphasise the
difference again: in the third approach, the free propagator is damped, whereas
our model possesses ordinary free propagators, but damped internal lines.

In Section \ref{sec3}, we will consider 1-loop corrections in order to extend the
classical theory.
We will see that these contributions are finite.

\section{\label{sec2} Smeared Field Operators}


We want to study the effect of replacing the scalar field operators $\phi(x)$ by
blurred operators, smeared over spacetime
\be
\label{a0}
\phi_M(x) \equiv N \int d^na\,\, e^{-a^T a} \,\phi(x+M a),
\ee
where $a$ is a real Euclidean n-dimensional vector, $M$ is a real $4\times n$
matrix. $N$ denotes a normalisation constant. The integration parameters $a^i$
are assumed to be dimensionless. Therefore, the matrix elements of $M$ have
dimension of length. The non-vanishing matrix $M$ generates the non-locality. We
will denote Minkowski indices by Greek letters, Euclidean indices by Roman
letters. Therefore, the index structure of $M$ is $M^\mu{}_i$. However, the case
$n>4$ can be reduced to the case $n=4$. Due to the QR-decomposition, the
matrix $M$ can be written as a product of the $4\times n$ matrix $\tilde R$
and an orthogonal $n\times n$ matrix $Q$. The first $4$ rows of $\tilde R$
contain a lower triangular $4\times 4$ matrix $R$, all other entries are zero,
\be
\label{aa}
M=\left[ \begin{array}{cc} R& 0  \end{array} \right] \, Q \equiv \tilde R \, Q.
\ee
The orthogonal matrix $Q$ can be absorbed in an integral transformation, $\tilde
a = Q \, a$, and we get
\be
\phi_M(x) = N \int d^n\tilde a\,\, e^{-\tilde a^T \tilde a} 
\,\phi(x+ \tilde R\,
 \tilde a).
\ee
Since $\tilde R$ has the form shown in (\ref{aa}), the integration over the
variables $\tilde a_5,\dots,\tilde a_n$ are Gau\ss ian integrals which merely
redefine the normalisation constant.
Hence, only $4$ dimensions are left. From now on, we will stick to
that case. 

Since the newly defined field operators $\phi_M(x)$ are superpositions of the
operators $\phi(x)$, we demand that they are solutions of the free Klein-Gordon
equation,
\be\label{a1}
\left( \Box_x + m^2 \right) \phi_M(x) = 0.
\ee
The Fourier transform is given by
\be\label{a2}
\phi ( x + M a ) = \int {d^4 k \over (2\pi)^2} \, e ^ { ik(x + Ma) } \tilde \phi 
(k).
\ee
Due to the Klein-Gordon equation, we can find a nice expression for the smeared
field operators $\phi_M(x)$,
\begin{eqnarray}
\phi_M(x) &=& (2\pi)^{-3/2} N \int \frac{d^3p}{\sqrt{2 \omega_p}}
  \left[
     b (\vec p) e^{-ip^+x} + b^\dagger(\vec p) e^{i p^+ x}
  \right] \nonumber\\&& \times
  \int d^4 a \,e^{-a^ra^r+ i p^+_\mu {M^\mu}_r a^r}\nonumber\\
\label{phi_M}
  &=& (2\pi)^{-3/2} \pi^2 N
  \int \frac{d^3p}{\sqrt{2 \omega_p}}
  \left[
     b (\vec p) e^{-ip^+x} + b^\dagger(\vec p) e^{i p^+ x}
  \right] \label{a3}\\&& \times\,
  \exp(-\frac14 p^+_\mu p^+_\nu \kappa^{\mu\nu}),\nonumber
\end{eqnarray}
where $p^+_\mu = (+\omega_p, -\vec{p})$ with $\omega_p = \sqrt { {\vec{p}}\,{}^2 +
m^2}$. $b$ and $b^\dagger$ obey the canonical commutation relations
$$
{}[b(\vec p),\, b^\dagger(\vec q)] = \delta^3 (\vec p - \vec q).
$$
Summation over repeated indices is implied. Furthermore, we have used 
the definition
\be
\kappa^{\mu\nu} \equiv {M^\mu}_r{M^\nu}_r=(MM^T)^{\mu\nu}.
\ee
The matrix $\kappa$ is symmetric. For real $M$, its eigenvalues are always 
bigger than or equal to zero, i.e. $\kappa$ is positive semidefinite. The
exponential factor in (\ref{a3}) is always damping,
$$
\exp(-\frac14 p^+_\mu p^+_\nu \kappa^{\mu\nu}) \le 1.
$$
As we will see below, $\kappa^{\mu\nu}$ 
characterises the perturbation theory, not $M$ itself. Therefore, we only have to
choose an appropriate matrix $\kappa^{\mu \nu}$ in order to do perturbation
theory, ensuring that the matrix can be reproduced by $MM^T$.
A tempting choice is $\kappa^{\mu \nu} \propto g^{\mu \nu}$, but $g$ is 
neither positive nor negative semidefinite. The choice $\kappa=0$ reproduces 
local field theory. 

We want to study the perturbative quantisation of this kind of deformation,
according to the results presented in \cite{Denk:2003jj}. The deformed
Hamiltonian is defined as
\begin{equation} \label{eq:H0pVs}
  H^* = H_0 + V^*,
\end{equation}
where $H_0$ denotes the free undeformed Hamiltonian of the theory. We have
replaced the scalar fields by the smeared fields (\ref{a0}), $\phi\to\phi_M$ 
in the interaction part of the Hamiltonian only. The free Hamiltonian, $H_0$ is
unaltered. Of course, it would be more natural to deform $H_0\rightarrow H_0^*$
also. Then the applicability of the perturbation theory elaborated in 
\cite{Denk:2003jj} is related to the question whether $H^*_0=H_0$ is true
or not. If $H^*_0\ne H_0$, we have to define the interaction Hamiltonian as
$\tilde V = V^* + \big( H^*_0-H_0 \big)$. In this case, we also have to make
sure that the time dependence of $\tilde V$ is given by
\be
\tilde V (t) = e^{i H_0 t} \, \tilde V(0) \, e^{-i H_0 t},
\ee
and the asymptotic behaviour is still governed by $H_0$ and not $H_0^*$.

Let us examine perturbation theory arising from Eq.~(\ref{eq:H0pVs}), leaving
the free Hamiltonian $H_0$ undeformed. 
The interaction corresponding to $\phi^k$ is deformed as follows:
\begin{eqnarray}
  V^*(x^0) & \equiv & {\lambda\over k!} \int d^3x \,\phi_M^k(x)\nonumber\\
\label{potential}  
  & = & {\lambda\over k!} \,N^k \int d^3x\int d^4 a_1 \ldots d^4 a_k \\
&&  \times \, e^{-\sum_i a_i^T a_i}\,
  \phi(x+Ma_1)\ldots\phi(x+Ma_k). \nonumber 
\end{eqnarray}
This is obviously translation invariant. Therefore, we will first relate Eq.
(\ref{potential}) to the notation introduced in \cite{Denk:2003jj} in order to
apply the momentum space rules given there, for a general non-local interaction.
The interaction has the general form
\be
V(z^0) = \int d^3 z \int d\underline{\mu} \,\, w(\underline{\mu}) \,
\phi(z+h_1(\underline{\mu})) \cdot \dots \cdot \phi(z+h_k(\underline{\mu})),
\ee
where
\bea
\nonumber
\underline{\mu} & = & (a^1_1, a^2_1, a^3_1, a^4_1, a^1_2,\dots a^3_k, a^4_k), \\
\nonumber
w(\underline{\mu}) & = & e^{-\sum_ {j=1}^k a^T_j a_j}\\
\nonumber
h_s(\mu) & = & M\cdot a_s, \, s=1,\dots, k. 
\eea
Following the procedure presented in \cite{Denk:2003jj}, we obtain the Feynman 
rules evaluating the Gell-Mann-Low formula
\bea
\label{gell-mann-low}
 \langle 0 | T \phi(x_1) \dots \phi(x_k) | 0 \rangle_H & = & \sum_{m=0}^\infty { (-1)^m \over m!}
        \int_{-\infty}^\infty dt_1 \dots \int_{-\infty}^\infty dt_m \\ 
\nonumber & \times &
        \langle 0 | T \phi(x_1) \dots \phi(x_k) V^*(t_1) \dots V^*(t_m) | 0
	\rangle _{(0)},
\eea
with the interaction potential~(\ref{potential}). 
In the above formula, $H$ indicates the Heisenberg picture and $(0)$ the fact 
that we use free fields of the Dirac picture on the r.h.s. For simplicity, 
we will drop the index $(0)$.
It is important to note that time ordering is performed with
respect to $x_1^0,\dots, x_k^0, t_1, \dots t_m$. The time arguments within a vertex (cf. Eq.~(\ref{potential}))
are not dissolved (TOPT) \cite{Liao:2002pj,Bahns:2002vm,Bozkaya:2002at,
Fischer:2003jh}.

The first step is to draw all possible momentum space diagrams with $k$ external legs, as 
described in \cite{Denk:2003jj}. We have to label each line with its 4-momentum $p_i$ 
including its direction and  the variable $\sigma_i$, where 
$p^\sigma_\mu =(\sigma \omega_{p}, -\vec{p})^T$, $\sigma=\pm 1$. 
To each line - with labels $p_i$
and $\sigma_i$ - we have to assign the factor
\be
\label{line}
{-i \over p_i^2+ m_i^2-i\epsilon} \,\, {\omega_{{p}_i}+\sigma_i p^0_i \over 
2\omega_{{p}_i}}.
\ee
The function $\chi$ is associated with each vertex:
$$
\chi(p_1^{\sigma_1},\ldots, p_k^{\sigma_k}) = {\lambda\over k!}
N^k  \int d^n a_1 \ldots d^n a_k \,
  e^{-\sum_j a_j^T a_j } \sum_{Q\in S^k}  
  \exp(-i \sum_j p^{\sigma_j}_j M \, a_{Q_j})
$$
\be
\label{eq:chi}
  = \lambda  \, \exp(-\frac14 \sum_i {p_i^{\sigma_i}}^T \kappa \, p^{\sigma_i}_i)
\ee
where we have summed over all permutations $Q\in S^k$ of the external momenta. 
By definition, the above integral is independent of the order of the momenta $p_i$.
Remarkably, there are only on-shell momenta involved because of Eq.~(\ref{phi_M}).
We have chosen
$$
  N = \pi^{-2}.
$$
Note that
\begin{equation}
p^T \kappa \, q = p_\mu q_\nu \, \kappa^{\mu\nu} 
\end{equation}
and
$$
  \kappa^{\mu\nu} = \kappa^{\nu\mu}.
$$
Additionally, we have to introduce the usual symmetry factor ${1\over S}$ and 
to assure momentum conservation at each vertex,
\be
(2\pi)^4 \delta^4(p_1 +\dots + p_k).
\ee
Finally, we have to integrate over all internal momenta $q_r$ which are not fixed by 
momentum conservation  
\be
\prod_{r=1}^{\#\mbox{\tiny Loops}} {d^4 q_r \over (2\pi)^4} 
\ee 
and sum over all $\sigma_i$'s.

As an example, let us consider the contribution of a line between two 
internal points belonging to different interaction regions ("internal
propagator"), i.e. corresponding
to different interaction potentials $V(t_i)$ in Eq.~(\ref{gell-mann-low}).
Therefore, we have to account for a line labelled by $q$ and $\sigma$ and two
vertices characterised by $\chi(q^\sigma,\dots)$ and $\chi(-q^\sigma,\dots)$,
respectively. Sticking everything together yields
\be
\label{int-prop}
\Delta_M (x-y) = {-i\over (2\pi)^4 } \int d^4 q \, { e ^ { -iq(x-y) } \over q^2 - m^2 
+ i\epsilon } 
\sum_{\sigma=\pm 1} {\omega_q + \sigma q^0 \over 2 \omega_q } e^ {- {1\over 2}
q^{\sigma T}  \kappa q^\sigma
}.
\ee
Eq.~(\ref{int-prop}) for the "internal propagator" can also be obtained by 
contracting two smeared field operators (\ref{phi_M}),
\be
\label{wsm}
\langle 0 | T \, \phi_M (x) \phi_M (y) | 0 \rangle = \Delta_M (x-y).
\ee
The time ordered product can easily be written as a sum of two terms
\bea
\label{t1}
\langle 0 | T \, \phi_M (x) \phi_M (y) | 0 \rangle & = & \langle 0 | \phi_M (x)
\phi_M (y) | 0 \rangle \, \theta(x^0 - y^0) \\
& + & \langle 0 | \phi_M (y) \phi_M (x) | 0 \rangle \, \theta(y^0 - x^0).
\nonumber
\eea
Inserting Eq.~(\ref{phi_M}) and the integral representation of the Heaviside
step function
\be
\label{heaviside}
\theta(t' - t) = \lim_{\epsilon\to 0} \, {-1 \over 2\pi i} \int_{-\infty}^\infty
{d \tau \over \tau + i\epsilon} \, e^{-i\tau (t'-t)} 
\ee
into (\ref{t1}) yields
\bea
&& \hspace{-1cm}
\lim_{\epsilon\to 0} \, {-1\over 2\pi i}
\int {d^3 k \, d \tau \over (2 \pi)^{3} \, 2 \omega_k} \,\, 
        e^{ - k^+_\mu k^+_\nu \, \kappa^{\mu\nu} / 2 } 
        \Bigg( 
	 e^{ -i\omega_k (x^0 - y^0) + i \vec{k} (\vec{x}-\vec{y}) }\, 
        { e^{i\tau(x^0 - y^0)}  \over \tau - i\epsilon }
\\
\nonumber
&& \hspace{3cm} +
e^{ i\omega_k (x^0 - y^0) - i \vec{k} (\vec{x}-\vec{y}) }\, 
	{ e^{i\tau(x^0 - y^0)} \over \tau + i\epsilon }
\Bigg).
\eea
The exponential damping is the only difference 
to the usual local calculation. After some substitutions and noting that
for the substitution $\vec{k}\to -\vec{k}$ we get $k^+\to -k^-$ we obtain the 
desired result:
\be
\label{www}
\langle 0 | T \, \phi_M (x) \phi_M (y) | 0 \rangle = 
{-i\over (2\pi)^4 } \int d^4 q \, { e ^ { -iq(x-y) } \over q^2 - m^2 
+ i\epsilon } 
\sum_{\sigma=\pm 1} {\omega_q + \sigma q^0 \over 2 \omega_q } e^ {- {1\over 2}
q^{\sigma T}  \kappa q^\sigma
}.
\ee
Eq. (\ref{wsm}) allows also a different interpretation for the Feynman rules.
Namely, we can attribute an exponential damping factor 
\be
e^{-{1\over 2} {q^\sigma}^T\, \kappa \, q ^\sigma }
\ee
to internal lines labelled by $q,\sigma$. The damping can be assigned either to
the internal lines or to the vertices. Of course, the amplitudes are unaffected 
by this choice.

In the situation
discussed here, free propagators are not changed, since
\be
G(p) = \sum_\sigma {-i \over p^2 - m^2 + i\epsilon} \,\, {\omega_p+\sigma p^0 
\over  2\omega_p}
= {-i \over p^2 - m^2 + i\epsilon}
\ee
with $p^0 = \omega_p$ for external particles.

In the next Section, we will examine 1-loop corrections and show that they are
all finite. Let us first discuss specific choices of the matrix $\kappa$,  
respectively $M$. For simplicity, we concentrate on the case of a diagonal
matrix $\kappa$.

The first choice we want to consider is the unit matrix,
\be
(\kappa^{\mu\nu}) = 2 \, \zeta \, \mathbf 1.
\ee
This can be accomplished, for example by using the following matrix $M$:
\be
\label{m1}
({M^\mu}_r) = \sqrt{ 2 \, \zeta } \,  \left(
\begin{array}{cccc} 0 & 1 && \\ -1 & 0 && \\ && 0 & 1 \\ && -1 & 0 \end{array}
        \right). 
\ee
The motivation to use an antisymmetric Matrix $M$ of full rank has already 
been stressed in \cite{Smailagic:2003rp}.  
We want to relate this approach to the non-commutativity of space-time. One of
the block diagonal matrices of (\ref{m1}) is related to the non-commutative
structure in \cite{Smailagic:2003rp}, cf. Eq. (\ref{b1}) with $\theta= 
{ 2 \, \zeta }$. 
Explicitly we have
\be
p^+_\mu \kappa^{\mu\nu} p^+_\nu = 2 \zeta ( \vec{p}\, ^2 + \omega_p^2) 
= 2 \zeta ( 2 \vec{p}\, ^2 + m^2 ),
\ee
where the second term can be absorbed within the normalisation constant in
(\ref{phi_M}). Therefore, this case is equivalent to choosing $\kappa^{00}=0$.
In general, the case $\kappa^{0i}=0$ is equivalent to the case 
$\kappa^{0\mu}=0$.

The smearing of the field operators considered in the next Section will only 
extend over the spatial dimensions, and the zero 
component of the 4-vector $Ma$ in (\ref{a0}) vanishes. In this case the Feynman
rules become simpler.
The factor $\chi$ associated to vertices becomes
\be
\chi(p_1^{\sigma_1},\ldots, p_k^{\sigma_k}) =
 \lambda  \, \exp(-\frac14 \sum_i \vec{p}_i\,^T \tilde\kappa \, \vec{p}_i),
\ee
which only contains the spatial components of the incoming momenta. We will
examine the cases
\bea
\nonumber
\tilde\kappa & = & 2\, \zeta 
        \left(\begin{array}{ccc} 1&&\\&1&\\&&1 \end{array}\right),\\
\label{kappa-def}
\tilde\kappa & = & 2\, \zeta 
        \left(\begin{array}{ccc} 1&&\\&1&\\&&0 \end{array}\right),\\
\nonumber
\tilde\kappa & = & 2\, \zeta 
        \left(\begin{array}{ccc} 1&&\\&0&\\&&0 \end{array}\right).
\eea
Therefore, the $\sigma_i$'s are only contained in the contributions assigned to lines, cf.
Eq.~(\ref{line}). They easily factorise, and
\be
\sum_{\sigma_1} {\omega_{q_1} + \sigma_1 q^0_1 \over 2 \omega_{q_1} } \ldots
\sum_{\sigma_k} {\omega_{q_k} + \sigma_k q^0_k \over 2 \omega_{q_k} } = 1. 
\ee
Hence, we have to assign to every line the usual factor
\be
{-i \over p_i^2 - m_i^2 + i\epsilon}.
\ee


\section{\label{sec3}Perturbative Corrections and\\ Power Counting}

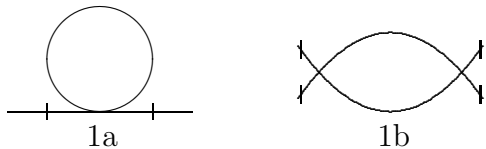
\begin{figure}[htp]
\begin{center}
\begin{picture}(200,50)(-10,0)

\put(50,30){\circle{40} }
\put(15,10){\line(1,0){70}}
\put(30,13){\line(0,-1){6}}
\put(70,13){\line(0,-1){6}}
\put(45,-3){1a}

\curve(125,15, 160,40, 195,15)
\curve(125,35, 160,10, 195,35)
\put(126,13){\line(0,1){7}}
\put(126,37){\line(0,-1){7}}
\put(194,13){\line(0,1){7}}
\put(194,37){\line(0,-1){7}}
\put(155,-3){1b}
\end{picture}
\end{center}
\caption{1-loop contributions for $\phi^4$ theory}
\label{fig1}
\end{figure}

In this Section, some properties of perturbative calculations with damped 
scalar field propagators will be studied. First, we will elaborate a power 
counting criterion by examining tadpole loops as shown in Figs.~\ref{fig1} and
\ref{fig3}. 
Finally, this criterion will be tested for various calculations in Euclidean 
as well as Minkowski space.


\vspace{1cm}

As indicated in the previous Section, the first $j\le4=d$ matrix elements in 
the diagonal of $\kappa$ are assumed to be 1, whereas all the other elements 
are assumed to be zero.
  
For simplicity, the damping factor will be kept track of by putting it into 
the damped "internal propagator"
\begin{equation}
\Delta_j(k) \equiv \frac{e^{-\zeta  \sum_{i=1}^j k_i^2}}{k^2 - m^2 + i \epsilon}
\end{equation}
as already indicated by Eq.~(\ref{int-prop}). $\zeta$ has dimension of $[length]^2$, possibly related to the deformation 
parameter of NCQFT \cite{Chaichian:1998kp,Cho:1999sg,Smailagic:2003rp}.
$j$ denotes the number of damped dimensions. 
The case $j=d$ actually does not fit into our approach of smeared field 
operators, since the zero component of the occurring momenta are never involved 
in the damping factor, cf. Eq. (\ref{phi_M}). Only on-shell momenta occur.
However, we will also treat this case in the Euclidean 
theory since it is manifestly covariant, and it does not make much extra work.
The definitions
\begin{eqnarray}
\bar k^2 & \equiv &  \sum_{i=1}^j k_i^2, \\
{k'}^2 & \equiv & \sum_{i=j+1}^d k_i^2
\end{eqnarray}
will also be helpful.


\subsection{\label{sec3.1}Power Counting}

In order to get a feeling for the power counting behaviour of perturbative
calculations with damped internal scalar field propagators in a $d$-dimensional
space-time, we present some general statements. Any vertex function is
characterised by the number of external lines $E$, the number of internal lines
$I$ and the number of interaction vertices $V$. 

A general vertex in coordinate space is of the form
\be
\label{s1}
V_i = \int d^d x\, \partial^{\delta_i}_x \phi^{b_i}(x),
\ee
where $\delta_i$ counts the number of derivatives, and $b_i$ stands for the
number of scalar fields involved in the interaction.

Let us first consider full damping, i.e. $j=d$.
The "internal propagators" described in Section \ref{sec2} are assumed to have
the following form in an Euclidean formulation
\be
\label{s2}
\int {d^d k \over (2\pi)^d} e^{ik(x-y)} {1\over k^2 + m^2} e^{-\zeta k^2},
\ee
neglecting some factors, which are not important for our considerations here, 
cf. Eq. (\ref{www}). In momentum space, this involves
\be
\label{s3}
\Delta_M (k) = \frac1{k^2+m^2} \, e^{-\zeta k^2},
\ee
where $k^2= (k^0)^2+\vec{k}\,{}^2$. For a fixed $n$, we rewrite (\ref{s3}) as
\bea
\label{s4}
\Delta_M^n (k) & = & \frac1{ (k^2+m^2) (1+\zeta k^2 + \dots + {1\over n!} (\zeta
	k^2)^n ) }\\
\nonumber
& = & \frac1{ (\zeta k^2)^n (k^2 + m^2)(1/n! + \mathcal O \left( 1/(\zeta k^2)^j
	\right) },
\eea
with $1\le j \le n$. In order to estimate the high momentum behaviour of 
$\Delta_M^n (k)$ it is sufficient to use 
\be
\label{s5}
\Delta_M^n (k) \approx {n! \over (\zeta k^2)^n k^2}.
\ee
For all high momenta $k$ there is a Polynomial $P_n(k^2)$ of degree $n\in\mathbb N$ such that 
$$
e^{\zeta k^2} > P_n(k^2) \mbox{ and } e^{-\zeta k^2} < \frac1{P_n(k^2)}.
$$

The superficial degree of divergence of any vertex graph $\gamma$ is therefore
given by
\be
\label{s6}
D_n(\gamma) = dL - (2n+2) I + \sum_{i=1}^V \delta_i.
\ee
Using
\be
\label{s7}
L = I - (V-1)
\ee
and the total number of all lines running to all vertices
\be
\label{s8}
\sum_i b_i = 2I + E,
\ee
we get for Eq. (\ref{s6})
\be
\label{s9}
D_n(\gamma) = d - \mbox{dim}\,\phi\, E - \sum_i (d-d_i) - 2n I.
\ee
The dimension of the scalar field 
is given by 
\be
\label{s10}
\mbox{dim}\,\phi = \frac{d}2 - 1,
\ee
and the corresponding dimension of the interaction vertex $V_i$ is defined as
\be
\label{s11}
d_i \equiv \delta_i + (\frac{d}2 - 1)\, b_i.
\ee
For $n=0$, we have the usual power counting. Now we are in the position to 
discuss specific models. 

In $d=3$ space-time dimensions, we have two classical interactions
\be
\label{s12}
V_1^3 = {\lambda_1 \over 4!} \int d^3 x\, \phi^4 (x) \,\, \mbox{ and } \,\,
V_2^3 = {\lambda_2 \over 6!} \int d^3 x\, \phi^6 (x).
\ee
In this case, dim$\phi=1/2$. This implies that $\lambda_1$ has dimension of a
mass, and $\lambda_2$  is dimensionless. The corresponding analogous
interaction of a $\phi^4$-model is (dim$\phi=1$)
\be
\label{s13}
V_3^4 = {\lambda_3 \over 4!} \int d^4 x\, \phi^4 (x).
\ee

\begin{figure}[htp]
\begin{center}
\begin{picture}(350,50)(-10,0)

\put(50,30){\circle{40} }
\put(15,10){\line(1,0){70}}
\put(30,13){\line(0,-1){6}}
\put(70,13){\line(0,-1){6}}
\put(45,-3){2a}

\put(145,10){\line(1,0){70}}
\put(180,30){\circle{40} }
\put(160,13){\line(0,-1){6}}
\put(200,13){\line(0,-1){6}}
\put(150,0){\line(3,1){30}}
\put(210,0){\line(-3,1){30}}
\put(155,4){\line(0,-1){6}}
\put(205,4){\line(0,-1){6}}
\put(175,-3){2b}

\curve(275,15, 310,40, 345,15)
\curve(275,35, 310,10, 345,35)
\put(270,22){\line(4,1){13}}
\put(270,28){\line(4,-1){12}}
\put(350,22){\line(-4,1){13}}
\put(350,28){\line(-4,-1){12}}
\put(276,14){\line(0,1){5}}
\put(276,36){\line(0,-1){5}}
\put(344,14){\line(0,1){5}}
\put(344,36){\line(0,-1){5}}
\put(271,21){\line(0,1){3}}
\put(349,21){\line(0,1){3}}
\put(271,29){\line(0,-1){3}}
\put(349,29){\line(0,-1){3}}
\put(305,-3){2c}
\end{picture}
\end{center}
\caption{1-loop graphs}
\label{fig3}
\end{figure}
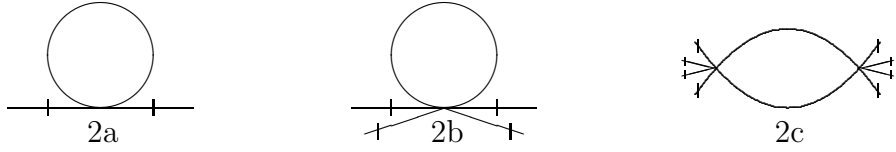

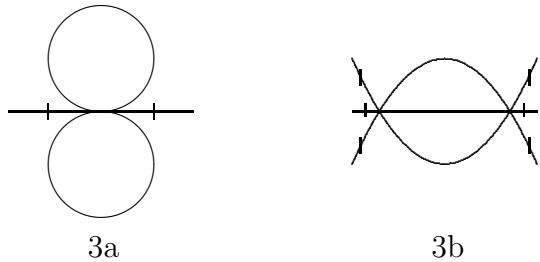
\begin{figure}[htp]
\begin{center}
\begin{picture}(210,100)

\put(50,80){\circle{40}}
\put(50,40){\circle{40}}
\put(15,60){\line(1,0){70}}
\put(30,63){\line(0,-1){6}}
\put(70,63){\line(0,-1){6}}
\put(45,5){3a}

\put(145,60){\line(1,0){70}}
\curve(145,80, 180,40, 215,80)
\curve(145,40, 180,80, 215,40)
\put(150,58){\line(0,1){5}}
\put(210,58){\line(0,1){5}}
\put(148,76){\line(0,-1){6}}
\put(148,44){\line(0,1){6}}
\put(212,76){\line(0,-1){6}}
\put(212,44){\line(0,1){6}}
\put(175,5){3b}

\end{picture}
\end{center}
\caption{2-loop corrections}
\label{fig4}
\end{figure}

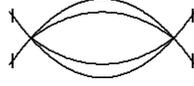
\begin{figure}[htp]
\begin{center}
\begin{picture}(100,50)

\curve(15,15, 50,40, 85,15)
\curve(15,35, 50,10, 85,35)
\put(16,14){\line(0,1){5}}
\put(16,36){\line(0,-1){5}}
\put(84,14){\line(0,1){5}}
\put(84,36){\line(0,-1){5}}
\curve(23,25, 50,35, 77,25)
\curve(23,25, 50,15, 77,25)

\end{picture}
\end{center}
\caption{3-loop correction}
\label{fig5}
\end{figure}

For $d=3$, some perturbative corrections up to third order are shown in
Figs.~\ref{fig3}-\ref{fig5}.


\vspace{1cm}


\vspace{1cm}


\vspace{1cm}

According to Eq. (\ref{s9}), we have the following degrees of divergence for
these classes of radiative corrections:

\vspace{1cm}
\begin{center}
\begin{picture}(350,60)
\put(0,50){ \line(1,0){400}}
\put(160,0){\line(0,1){70}}
\put(320,0){\line(0,1){70}}

\put(10,63){Fig. 2a}
\put(5,25){$D_n=1-2n <0 ,\, \forall n>0$}
\put(5,10){$n=0:D_0=1$}

\put(170,63){Fig. 2b}
\put(165,25){$D_n=1-2n <0 ,\, \forall n>0$}
\put(165,10){$n=0:D_0=1$}

\put(330,63){Fig. 2c}
\put(325,25){finite}
\end{picture}
\end{center}

\vspace{1cm}
\begin{center}
\begin{picture}(300,60)
\put(0,50){ \line(1,0){300}}
\put(160,0){\line(0,1){70}}

\put(10,63){Fig. 3a}
\put(5,25){$D_n=2-4n <0 ,\, \forall n>0$}
\put(5,10){$n=0:D_0=2$}

\put(170,63){Fig. 3b}
\put(165,25){$D_n=-6n <0 ,\, \forall n>0$}
\put(165,10){$n=0:D_0=0$}
\end{picture}
\end{center}

\vspace{1cm}
\begin{center}
\begin{picture}(150,60)
\put(0,50){ \line(1,0){150}}

\put(10,63){Fig. 4}
\put(5,25){$D_n=1-8n <0 ,\, \forall n>0$}
\put(5,10){$n=0:D_0=1$}
\end{picture}
\end{center}

For $d=4$, the corrections at the one-loop level are shown in
Fig.~\ref{fig1}. The degrees of
divergence are given by:

\vspace{1cm}
\begin{center}
\begin{picture}(300,60)
\put(0,50){ \line(1,0){300}}
\put(160,0){\line(0,1){70}}

\put(10,63){Fig. 1a}
\put(5,25){$D_n=2-2n <0 ,\, \forall n>1$}
\put(5,10){$n=0:D_0=2$}

\put(170,63){Fig. 1b}
\put(165,25){$D_n=-2n <0 ,\, \forall n>0$}
\put(165,10){$n=0:D_0=0$}
\end{picture}
\end{center}

Thus, we have finiteness for all above mentioned graphs with fully damped
propagators for $n>1$.

In order to describe the power counting behaviour of the tadpole contribution
with a partial damping in some directions in the Euclidean formulation, we have
to consider the following integral
\begin{equation}
  \Gamma_{tp}^j \equiv \frac1{(2\pi)^d}\int d^dk\,\, 
  \Delta_j (k)\label{eq:tp1}\,\, .
\end{equation}
This integral can be rewritten as 
\begin{equation}
\label{sB}
\Gamma_{tp}^j =\frac1{(2\pi)^d}
	\int d^j\bar k \,\,e^{-\zeta \, \bar k^2} \int d^{d-j}k' \,\,
	\frac1{  \bar k^2 + {k'}^2 + m^2}\,\, .
\end{equation}
The case $j=d$ has already been discussed. Now, we approximate Eq. (\ref{sB})
with a finite parameter $l$ as
\be
\label{sC}
\Gamma_{tp}^{j,l} = \frac1{(2\pi)^d} \int d^j\bar k\,\,
	\frac1{ 1+\dots + {1\over l!} (\zeta \bar k^2)^l }\,
	\int d^{d-j}k' \,\,\frac1{  \bar k^2 + {k'}^2 + m^2}\,\, .
\ee
For $0 < j \le d$, there exists always a $l>0$ such that the $\bar k$-integration
converges.
 It remains to estimate the 
$k'$ integration. Naive power counting can be applied. For the tadpole, we get
\begin{equation}
\label{sE}
D_j=(d-j)-2,
\end{equation}
For $d=3$ and $j=2$, one has $D_j=-1$. This will be checked by explicit
calculations in Section~\ref{sec3.1}.

For $d=4$, we conclude from (\ref{sE}) that the degree of
damping has to be $j>2$ in order to have convergence. We will see that these
results are compatible with direct calculations.

Using the same philosophy, we can discuss an arbitrary $L$-loop contribution. We
can estimate the naive power counting (assuming that the integration 
over the $j$ damped directions is convergent) by
\begin{equation}
\label{sF}
D_j=L(d-j)-2I + \sum_i \delta_i ,
\end{equation}
implying
\be
\label{sG}
D_j = d -\mbox{dim}\phi \, E - \sum_i (d-d_i) - jL.
\ee
Eq. (\ref{sG}) seems to imply that the superficial degree of divergence $D_j$
linearly decreses with the number of loops $L$. But $L$ and the number of
vertices are related. We can rewrite Eq. (\ref{sG}) in the following way
\be
\label{w1}
D_j = d -j - E\, \mbox{dim}\phi + {Ej\over 2} - \sum_i \left( d - \delta_i
-b_i(\frac{d}{2}-1) \right) - \sum_i \left( {b_i\over 2} - 1 \right) j.
\ee
We see that $D_j$ decreases with the number of vertices and may increase 
with the number of external legs.

For $j=0$ (no damping), we get back the power counting behaviour of a local 
theory. 

As a further consistency check, we discuss Eq. (\ref{sG}) for the tadpole
contribution with $L=1$. For $d=3$, we have the following: 
$E=2$ and $d-d_i=1$, for the $\phi^4$ interaction; $E=4$ and $d-d_i=0$, for the
$\phi^6$ interaction. Therefore, both cases yield
\be
\label{sH}
D_j = 1-j.
\ee
This implies convergence for $j>1$.

For $d=4$, we find
\be
\label{sI}
D_j = 2-j,
\ee
meaning that convergence implies $j>2$. The fact that the degree of divergence
depends on the number of smeared dimension has also been observed in
\cite{Nicolini:2004yb} where the vacuum energy density has been discussed in the
framework of the third approach of Section \ref{sec1}.


\subsection{\label{sec3.2}Explicit Calculations in the Euclidean Case}

Let us consider the tadpole integral
\be
\Gamma_{tp}^j =(2\pi)^{-d}
\int d^dk \,\, \frac{e^{-\zeta \bar k^2}}{k^2+m^2}
\ee
in $d=3$ and $4$ dimensions.
In $3$ space-time dimensions, we have to solve the following integral
\be
\label{w2}
\Gamma^2_{tp} = (2\pi)^{-3}\int d^3k \,\, \frac{e^{-\zeta \bar k^2}}{k^2+m^2}.
\ee
The relevant loop graphs are Fig.~\ref{fig3}a and Fig.~\ref{fig4}a. We
employ the Schwinger parametrisation
\be
\label{w3}
\frac1{k^2+m^2} = \int_0 ^\infty d\alpha e^{-\alpha (k^2 + m^2)}
\ee
to obtain
\bea
\label{w4}
\Gamma^2_{tp} & = & {\pi^{3/2}\over (2\pi)^3} \int_0^\infty \frac1{\alpha^{1/2}
	(\alpha + \zeta)} e^{-\alpha m^2}\\
	\nonumber
& = & {\pi^{3/2}\over (2\pi)^3 }\,  e^{\zeta \,m^2} \sqrt{\pi\over \zeta}\,\,
	\Gamma(1/2, \zeta \, m^2).
\eea
$\Gamma(z)$ is the ordinary Gamma-function, whereas $\Gamma(1/2,z^2)$ denotes
the "finite" incomplete Gamma-function,
\be
\Gamma(1/2, \zeta\, m^2) = \sqrt{\pi} - \sum_{n=0}^ \infty {(-1)^n \over n!\,
(n+1/2)} (\zeta \, m^2)^{n + 1/2}. 
\ee
For $j=1$, the tadpole contribution diverges.

In $4$ dimensions, the calculations are a bit more involved. The tadpole
integral is given by
\be
\label{rt}
\Gamma^j_{tp} = (2\pi)^{-4}\int d^4k \,\, \frac{e^{-\zeta \bar k^2}}{k^2+m^2}.
\ee
But already at this step, it is clear that UV/IR mixing as it occurs in the first
approach of Section \ref{sec1} does not appear for the tadpole here. 
UV/IR mixing basically means that divergences due to UV-integrations arise for
vanishing external momenta. But here, the integration does not even depend on the external momenta.
Using the Schwinger parametrisation and carrying out the Gaussian integration, 
we obtain
\begin{equation}
  \Gamma_{tp}^j = \frac{\pi^2}{(2\pi)^4}\int_0^\infty d\alpha\,\, 
  \frac{e^{-\alpha m^2}}{\alpha^{2-j/2}(\alpha+\zeta)^j/2}.
\end{equation}
The possible problems of UV-integration are now hidden in the behaviour of this
integral for $\alpha\rightarrow 0$. The power counting behaviour can be studied
by regulating this expression which is done by restricting the integration  to 
$\alpha \in [1/\Lambda^2,\infty[$. By dividing this area of integration into
$[1/\Lambda^2,a[$ and $[a,\infty[$ with $a\ll \zeta$, we can read off the degree of 
divergence to be
\be
  D_j=2-j.
\ee
This agrees with the power counting formula given above and tells us that the 
tadpole is quadratically, linearly and logarithmically divergent for $j=0,1,2$, 
respectively. The minimum damping rendering the tadpole contribution finite is
given by $j=3$. We obtain
\begin{equation}
\Gamma_{tp}^3 = \frac{2 \pi^{5/2}}{(2\pi)^4\zeta} U(\frac12,0,m^2\zeta),
\end{equation}
where $U$ denotes the confluent hypergeomtric function, with $U(\frac12,0,0)=\frac2{\sqrt{\pi}}$.
Of course, also the case $j=4$ gives a finite result:
\begin{equation}
  \Gamma_{tp}^4 = \frac{\pi^2}{(2\pi)^4} \left[
    \zeta^{-1}+m^2 e^{\zeta m^2} {\rm Ei}(-\zeta m^2)
\right],
\end{equation}
where ${\rm Ei}$ is the exponential integral function with the following expansion for $x<0$:
\begin{equation}
  {\rm Ei}(x)= e +\ln(-x)+\sum_{k=1}^\infty \frac{x^k}{k (k!)},
\end{equation}
where $e$ is the Euler-constant.
Note that the parameter $\zeta$ acts as a regulator. For any $j$, the 
 integral (\ref{rt}) diverges quadratically for vanishing $\zeta$.

In a second step, we consider the one loop 4-point 1PI-vertex 
(see Fig.~\ref{fig1}b). The corresponding Feynman integral is
\begin{equation}
  \Gamma_{4}^j(p) \equiv \frac1{(2\pi)^4}\int d^4k\,\, 
  \Delta_j (k)\Delta_j (k+p)\label{eq:loop1}.
\end{equation}
Using two Schwinger parameters $\alpha$, $\beta$ and applying the integral 
transformation
\begin{eqnarray}
  \alpha &=& (1-\xi) \lambda,\\
\beta&=& \xi \lambda,
\end{eqnarray}
we get 
\begin{equation}\label{eq:EG4jfin}
  \Gamma_{4}^j(p) = \frac{\pi^2}{(2\pi)^4}
  \int_0^\infty d\lambda\int_0^1d\xi\,\,
  \frac{
    e^{-\sum_{i=1}^j p_i^2\frac{\xi(1-\xi)\lambda^2+\zeta(\lambda+\zeta)}{\lambda+2\zeta}
  -(p^2-p^2_i)\xi(1-\xi) \lambda -\lambda m^2
 }}
{\lambda^{1-j/2}(\lambda+2\zeta)^{j/2}}.
\end{equation}
A further evaluation of these integral is quite tricky. But the UV-behaviour can again be read off from the properties of the denominator 
$$
\lambda^{1-j/2}(\lambda+2\zeta)^{j/2}
$$
for $\lambda \rightarrow 0$. The only problems might arise from the first 
factor and we do not expect UV-divergences for $1-j/2<1$ or $j>0$. This means 
that at least one direction of space-time has to be damped in order to render the
integral 
$\Gamma_{4}^j$ finite, which again agrees with our power counting criterion $D_j=-j<0$. A more detailed analysis of the integral was only 
possible for $j=4$, where it could be rewritten after an appropriate 
transformation as
\begin{equation}
  \Gamma_{4}^j(p) = -\frac{\pi^2}{(2\pi)^4} e^{2 \zeta m^2}
\int_0^1 d\xi \, {\rm Ei}(-2[\xi(1-\xi)p^2+m^2]\zeta).
\end{equation}
This expression is finite since ${\rm Ei}$ in the integrand is evaluated at negative 
values only, where it is well behaved, and the integral itself is over a 
finite interval.


\subsection{\label{sec3.3}Explicit Calculations in the Minkowski Case}

Now we are ready to carry out a similar analysis for Minkowski space. 
The tadpole diagram  corresponds to the integral
\begin{equation}
\Gamma_{tp}^j \equiv \int d^4k\,\,\frac{e^{-\zeta \bar k^2}}
{k^2 - m^2 + i\epsilon}.
\end{equation}
The case of full damping 
(in all space-time directions) is omitted for Minkowski space, where we would 
have to use $\exp[-\zeta(k_1^2+k_2^2+k_3^2+k_0^2)]$ as a damping factor to ensure
finiteness. Wick rotation is not possible for the fully damped Minkowski 
situation since one would encounter exploding factors 
$\exp(-\zeta k_0^2)\rightarrow \exp(\zeta k_4^2)$. For the following 
discussion, the exponential is assumed not to depend on $k_0$. Hence, there 
are no obstacles opposing Wick rotation, and the results of the preceding 
discussion in Euclidean space for $j<4$ are directly applicable. 

We now turn to the more complicated kind of loops as shown in 
Fig~(\ref{fig1}).
The interesting part of this diagram is given by the integral
\begin{eqnarray}
\Gamma_{4}^j(p=p_1+p_2) &\equiv&
\int d^4k \,\,\frac{e^{-\zeta \bar k^2}}{k^2 - m^2 + i\epsilon}\,
\frac{e^{-\zeta \sum_i (k+p)_i^2}}{(k+p)^2 - m^2 + i\epsilon}\\
&\equiv& \int d^4k\,\,f(\vec k,\vec p) \,g(k,p), \nonumber
\end{eqnarray}
where
$$
f(\vec k,\vec p) \equiv e^{-\zeta \bar k^2} e^{-\zeta \sum_i (k+p)_i^2}.
$$
$f$ only depends on spatial momenta and not on their time component. The direct 
evaluation of $\Gamma^j_{4}$ for arbitrary external momenta $p$ seems to be 
rather tricky, and here we restrict ourselves to the UV-behaviour 
concerning the $k$ integration. We want to give an upper bound for $\Gamma^j_{4}$ 
and show that it is finite. But let us first get rid of the poles concerning 
the $k^0$ integration. This is most easily accomplished by the residue theorem
\begin{equation}
  I^0(\vec k,p)\equiv \int dk^0 \,\,g(k,p) =
  \pi i 
  \frac{\left(\frac1 {\qom k} + \frac 1 { \omega_{\vec k+\vec p}}\right)}
    {(\qom k+\omega_{\vec k+\vec p})^2-{p^0}^2}.
\end{equation}
The loop integral then reads
$$
\Gamma_{4}^j(p) = \int d^3k \,\,f(\vec k,\vec p)\, I^0(\vec k,p) .
$$
$I^0$ has the following bound
$$
  |I^0(\vec k,p)| \leq \frac C{|\vec k|^3} 
  \quad \quad {\rm for}\quad |\vec k|\geq r_p,
$$
where $C >0$ is some proportionality constant and 
$$
  r_p \propto \max(|p_0|,|\vec p|).
$$
Defining the UV-part of the integration 
$UV\equiv\{\vec k\in R^3||\vec k|\geq r_p\}$, we thus conclude
\begin{eqnarray}
|I^{UV}(p)| &\equiv& |\int_{UV} d^3k \,\,f(\vec k,\vec p)\, I^0(\vec k,p)|
\leq\int_{UV} d^3k \,\,|f(\vec k,\vec p)\, I^0(\vec k,p)|\nonumber
\\&\leq&\int_{UV} d^3k \,\,f(\vec k,\vec p)\,\frac C{|\vec k|^3} .
\end{eqnarray}
This is finite as long as the sum over $i$ within $f$ involves at least one of 
the three spatial components, say $j>0$. This is consistent with the results of the 
Euclidean discussion, where we concluded the same UV-behaviour by inspection of 
Eq.~(\ref{eq:EG4jfin}) for $\lambda\rightarrow 0$. It again confirms our power counting criterion.


\section{Conclusion and Remarks}

We have discussed a non-local real scalar field theory. The non-locality is
located in the interaction where we have replaced the usual local fields by
smeared field operators (\ref{a0}). The Feynman rules are worked out in 
Section~\ref{sec2} using the Gell-Mann-Low formula (\ref{gell-mann-low}). 
The free theory is not modified. Therefore,
also the free propagators are unaltered. As a result of the smearing, the vertex
contribution is exponentially damped by the incoming on-shell momenta 
(\ref{eq:chi}). The fact that on-shell momenta enter the vertex contribution is
of vital importance and a natural consequence of TOPT. In
contrast to this result, the exponentially damped propagators obtained in
\cite{Chaichian:1998kp,Cho:1999sg,Smailagic:2003rp} contain arbitrary 
momenta. 

In Section \ref{sec3}, we have carefully discussed UV properties of the model.
We have derived a power counting formula (55) which provides the superficial
degree of divergence for theories with exponential damping in arbitrarily many 
dimensions.
Explicit calculations of 1-loop diagrams in the Euclidean and Minkowski
framework, done in Sections~\ref{sec3.2} and resp.~\ref{sec3.3}, agree with the
result from the generalised power counting formula. In $d=3$ space-time
dimensions, the tadpole contribution shown in Fig.~\ref{fig3} is finite if at 
least one dimension is damped, i.e. $j > 1$. The other loop contribution 
in Fig.~\ref{fig1}a is finite independently of $j$. In $d=4$ space-time 
dimensions, the tadpole contribution converges for $j > 2$ and 
the 1-PI graph of Fig.~\ref{fig1}b 
for $j\ge 1$. The power counting formula shows that the presented model is UV 
finite to all orders in perturbation theory according to the proposed power
counting formula.
Notably, there is also no UV/IR mixing present at the 1-loop level.

Applying the methods presented here to gauge theories is the next interesting 
step and may provide new insights.

\subsection*{Acknowledgement}

This work has been supported by DOC (predoc program of the \"Oster\-rei\-chische 
Akademie der Wissenschaften) (S.D.) and by Fonds zur F\"orderung der
wissenschaft\-lichen Forschung (Austrian Science Fund), projects
P15015-N08~(V.P.) and P15463-N08~(M.W.).

\end{document}